\documentclass[conference]{IEEEtran}
\usepackage{cite}
\usepackage{multirow}
\usepackage{url}
\usepackage{amsmath,amssymb,amsfonts}
\usepackage{algorithmic}
\usepackage{comment}
\usepackage{graphicx}
\usepackage{textcomp}
\usepackage{xcolor}

\def\BibTeX{{\rm B\kern-.05em{\sc i\kern-.025em b}\kern-.08em
    T\kern-.1667em\lower.7ex\hbox{E}\kern-.125emX}}
\begin{document}

\title{Implementation of Ethereum Accounts and Transactions on Embedded IoT Devices}

\author{\IEEEauthorblockN{Giulia Rafaiani, Paolo Santini, Marco Baldi, Franco Chiaraluce} \IEEEauthorblockA{\textit{Department of Information Engineering} \\ \textit{Polytechnic University of Marche}, Ancona, Italy \\ \texttt{\{g.rafaiani, p.santini, m.baldi, f.chiaraluce\}@univpm.it}}}

\maketitle

\begin{abstract}
The growing interest in Internet of Things (IoT) and Industrial IoT (IIoT) poses the challenge of finding robust solutions for the certification and notarization of data produced and collected by embedded devices.
The blockchain and distributed ledger technologies represent a promising solution to address these issues, but rise other questions, for example regarding their practical feasibility.
In fact, IoT devices have limited resources and, consequently, may not be able to easily perform all the operations required to participate in a blockchain.
In this paper we propose a minimal architecture to allow IoT devices performing data certification and notarization on the Ethereum blockchain.
We develop a hardware-software platform through which a lightweight device (e.g., an IoT sensor), holding a secret key and the associated public address, produces signed transactions, which are then submitted to the blockchain network. 
This guarantees data integrity and authenticity and, on the other hand, minimizes the computational burden on the lightweight device. 
To show the practicality of the proposed approach, we report and discuss the results of benchmarks performed on ARM Cortex-M4 hardware architectures, sending transactions over the Ropsten testnet.
Our results show that all the necessary operations can be performed with small latency, thus proving that an IoT device can directly interact with the blockchain, without apparent bottlenecks.
\end{abstract}

\section{Introduction}
The constant progresses in communication technologies and miniaturization of electronic components and devices are leading to the integration of pervasive networking devices in our everyday life and enabling the so-called ``Internet of Things'' (IoT) paradigm. 
Indeed, the ultimate aim of the IoT is to provide physical objects with a digital identity and to create a digital twin of the real world. 
To do so, objects are combined with sensors, actuators, processors and communication software in order to make them able to interact and communicate with the external environment \cite{madakam}. 
IoT devices are, essentially, embedded systems connected to the Internet. 
In fact, an embedded system is a computer system built upon a microcontroller (roughly, a central processor unit (CPU) with memory and some peripherals), which additionally includes inputs/outputs and communication interfaces \cite{xiao}. 
Usually, embedded systems have small size, low cost and low power consumption; for these reasons, they are used in many devices we use everyday, such as domestic appliances, vehicles, smartwatches, and so on.
In fact, the low computational and storage capabilities of IoT devices clash with the need of finding reliable and sustainable ways to notarize and certificate the data that they continuously generate and exchange.
This leads to the necessity of introducing some supporting technology \cite{diaz};
a possible solution is to rely on the \textit{blockchain} technology. Basically, a blockchain is a ledger distributed among the participants of a peer-to-peer network.
Data that are stored on the ledger are approved by the majority of network nodes, that for this purpose implement a \textit{consensus protocol}.
This guarantees availability and immutability of data.
Due to these peculiarities, the blockchain technology is largely recognized as a promising solution to address reliability, privacy and security issues of IoT applications \cite{malviya}.
Each blockchain network is characterized by the underlying consensus protocol, to which network nodes must adhere.
There exists a wide and heterogeneous plethora of consensus protocols, each with its own set of rules and actions demanded to the users. 
Fully-fledged nodes are required to, at least, download and store large portions of the blockchain, and this consequently comes with a significant cost. 
Alternatively, a user may choose to interact with a blockchain in a lighter way, by only sending and receiving transactions (\textit{i.e.}, the basic instructions to make the blockchain evolve). 
Yet, even these lighter nodes need to execute functionalities such as hash functions, digital signatures obtained, for example, through the Elliptic Curve Digital Signature Algorithm (ECDSA), and data encoding, required to format transactions.
One may wonder whether such operations can be performed by IoT devices, given their limited resources, and, if so, with which cost.

We propose a microcontroller implementation of Ethereum-compliant accounts and transactions.
The target platform we consider is the ARM Cortex M-series, which is ideal for IoT applications because of its balance between low power consumption and high efficiency and performance \cite{yiu}.
Specifically, we consider the family of STM32 microcontrollers, which integrate the ARM Cortex-M4 processor.
The target blockchain is Ethereum, because of its large popularity and deployment.
Furthermore, Ethereum is at the basis of other blockchain and distributed ledger solutions, such as Hyperledger Besu: our implementation can be used, with only eventual and slight modifications, to interact with all Ethereum-compliant technologies.
Our solution has been validated with a complete test in which raw transactions are sent to Ropsten, a well-known Ethereum testnet. Our results show that the considered device can produce transactions in a time that is comparable to that required by the network to evaluate and accept them. Hence, our solution enables efficient interaction with the blockchain, without  performance nor latency issues.

The paper is organized as follows. In Section \ref{background}, we describe previous works and recall some background notions.
In Section \ref{thecost}, we provide quantitative arguments concerning the limited blockchain functionalities that can be implemented on a typical IoT device. 
In Section \ref{performances}, we describe our implementation and present the resulting benchmark results.
Finally, in Section \ref{conclusion} we draw some conclusions.

\section{Background}
\label{background}
In this section we provide a brief overview of related works and provide some background notions concerning the Ethereum blockchain.

\subsection{Related works}

Many previous works deal with the integration of IoT and blockchain \cite{reyna,conoscenti}. However, to the best of our knowledge, an analysis of this type, focused on embedded hardware platforms, has not appeared in the literature to date.
In \cite{ledwaba}, the authors evaluate the performance of ECDSA cryptographic functions on ARM Cortex-M microprocessors. These functions are also at the basis of Ethereum transactions, but no aspects related to the blockchain are considered in that work.
The authors of \cite{umran}, instead, study how IoT and blockchain can be integrated by proposing a blockchain-based IoT network, where some network nodes are based on ARM Cortex-M microprocessors; however, they propose a strictly customized blockchain, which works in a completely different way from Ethereum.
An analogous approach is proposed in \cite{latif}.
The project in \cite{firefly} contains the code necessary to build an Ethereum hardware wallet. However, the creation of the transaction is delegated to an external device and, moreover, the whole project is tailored to Arduino boards.

\subsection{Ethereum blockchain}

The blockchain literally is a chain of blocks, connected one another through cryptographic functions. Every block contains a list of transactions, \textit{i.e.}, cryptographically-signed data entries \cite{ethereum}. 
The blockchain is a distributed and decentralized ledger collecting these transactions, and every node in the network owns an identical copy of this register. The rules for appending a new block to the chain are dictated by the consensus protocol.
Ethereum is one of the best known blockchain infrastructures and its cryptocurrency, Ether, is second only to Bitcoin.
In order to send or receive a transaction, an Ethereum user should possess an account, that is, an entity with an associated 
private-public key pair and an address. When a user wants to send data or money to another user, they should digitally sign the transaction with their own private key and should know the recipient's address.
Then, the transaction has to be validated by the network to avoid any type of frauds,  before being immutably written onto the ledger.
Every Ethereum transaction contains the following fields:

\begin{itemize}
    \item \texttt{nonce} - the total number of transactions sent by the address that is sending the current transaction;
    \item \texttt{gasPrice} - the amount of money that should be paid per gas unit, that is, the fundamental network cost unit;
    \item \texttt{gasLimit} - the maximum gas amount that should be used for processing the current transaction;
    \item \texttt{to} - the address of the transaction recipient;
    \item \texttt{value} - the amount of money that should be transferred to the transaction recipient;
    \item \texttt{r,s,v} - the values corresponding to the transaction signature;
    \item \texttt{data} - the input data of the transaction.
\end{itemize}

All the parameters of the transaction, except for the \texttt{r}, \texttt{s}, \texttt{v} values, are encoded through RLP (Recursive Length Prefix) and the result is hashed through the function Keccak256. The obtained digest is then signed by the transaction sender, and the transaction becomes ready to be accepted by the network and included in a block.

\subsection{Elliptic Curve Digital Signature Algorithm}

ECDSA is a digital signature algorithm based on elliptic curves over finite fields.
A set of ECDSA domain parameters is a sextuple $(p,a,b,G,n,h)$, whose meaning is briefly recalled next. A prime integer $p$ is used to specify the finite ring $\mathbb{Z}_p = \mathbb Z\setminus p\mathbb Z$,  while
$a, b \in \mathbb{Z}_p$ are the parameters of the elliptic curve, defined as
\begin{equation}
\label{eq:elliptic_curve}
E : y^2 = x^3+ax+b\mod p.
\end{equation}
The elliptic curve $E(\mathbb Z_p)$ is defined as the set of all points $(x,y)\in\mathbb Z_p^2$ for which \eqref{eq:elliptic_curve} is satisfied.
The point $G\in E(\mathbb Z_p)$ is called base point, $n$ is a prime number specifying the order of $G$ and $h$ is the integer giving the cofactor of the curve \cite{ecdsa}.

An ECDSA key pair is given by $(d,Q)$, where $d$ is an integer in $[1 , n-1]$ and $Q = d\times G\in E(\mathbb Z_p)$, where $\times$ denotes the elliptic curve scalar multiplication (that is, the repeated addiction of the point $G$ along the curve). 
The security of the public key relies on the fact that, given $Q$ and $G$, to find the secret $d$ an attacker must solve the elliptic curve discrete logarithm problem, which is a well known hard problem.
To produce and verify  signatures, one has to perform basic arithmetic operations over $\mathbb Z\setminus n \mathbb Z$, as well as points additions and multiplications over the curve $E$.
We omit further details about these algorithms (which can be easily found in textbooks by the interested reader), and continue by recalling how ECDSA is employed in Ethereum.

\subsubsection*{ECDSA in Ethereum}
Ethereum utilises ECDSA with the curve \textit{secp256k1}, specified by the following parameters (expressed as hexadecimal numbers):

\begin{itemize}
    \item $p =$ FFFFFFFF FFFFFFFF FFFFFFFF FFFFFFFF FFFFFFFF FFFFFFFF FFFFFFFE FFFFFC2F;
    \item $a=$ 00000000 00000000 00000000 00000000 00000000 00000000 00000000 00000000;
    \item $b=$ 00000000 00000000 00000000 00000000 00000000 00000000 00000000 00000007;
    \item $G=$ 04 79BE667E F9DCBBAC 55A06295 CE870B07 029BFCDB 2DCE28D9 59F2815B 16F81798 483ADA77 26A3C465 5DA4FBFC 0E1108A8 FD17B448 A6855419 9C47D08F FB10D4B8;
    \item $n=$ FFFFFFFF FFFFFFFF FFFFFFFF FFFFFFFE BAAEDCE6 AF48A03B BFD25E8C D0364141;
    \item $h=$ 01.
\end{itemize}

Therefore, the resulting equation of the elliptic curve \textit{secp256k1} is $E: y^2=x^3+7$, where $p = 2^{256}-2^{32}-2^{9}-2^{8}-2^{7}-2^{6}-2^{4}-1$.

The private key is represented as a 32-byte array, while the public key is a 64-byte array. 

The Ethereum address is obtained as the last 160-bits of the Keccak256 hash of the public key. 
Inside the transaction, the signing parameters are given as the triplet $(\texttt{r,s,v})$. 
The $(\texttt{r,s})$ pair is computed as the result of the ECDSA signing algorithm.
The $\texttt{v}$ parameter is the \textit{recovery identifier}, \textit{i.e.}, a 1 byte value which is used to recover the sender public key from the pair $(\texttt{r}, \texttt{s})$.
In fact, considering an elliptic curve, for each value of $x$ there are two different possible values of $y$ such that $(x,y)\in E$. The parameter $\texttt{v}$ is used to discriminate between these two possible values: recovering the sender public key allows to both verify the signature and the address, which is not included among the transaction parameters.
The values that can be assumed by the parameter $\texttt{v}$ are different according to how the transaction is encoded (for more details about this, we refer the interested reader to  \cite{ethereum,eip155}).

\section{Choice of the Ethereum node type}
\label{thecost}
Ethereum nodes can be grouped into three main classes:
\begin{itemize}
\item[-] \textbf{full nodes}: download and store all the blocks. They actively participate in the consensus, proposing and validating new blocks.
Note that full nodes should actually be distinguished from the so-called \textit{archive nodes} which, when syncing to the chain, additionally compute and store all of the intermediate states (we can think of the state as an instantaneous picture of the blockchain overall situation).
Clearly an archive node needs a storage capability which is way larger than that of a full node.
As we argue in the following, even ordinary full nodes require hardware capabilities which are out of range for IoT devices, so we omit archive nodes from our discussion;
\item[-] \textbf{light nodes}: download and store only the block headers, and request every other information they need. They cannot propose and validate new blocks, and can only verify inclusion of transactions in accepted blocks;
\item[-] \textbf{accounts}: are associated to a cryptographic key pair and an address.
An account does not participate in the consensus and cannot verify transactions, but can only propose transactions.
\end{itemize}
For further details about the behaviour and properties of each of the considered node types, we refer the interested reader to the official Ethereum documentation.\footnote{\url{https://ethereum.org/en/developers/docs/nodes-and-clients/}}

Let us consider the burden that each type of node has to face, in terms of required storage resources.
To this end, we make use of the statistics offered by Etherscan,\footnote{\url{https://etherscan.io/charts}} and consider the situation of April 1st, 2022.
The blockchain was composed of 14,497,082 blocks, with an average block size of 87,026 B.
A new block was produced, on average, every 13.2 s.
The header size changes continuously, since it depends on some variables which do not have a fixed length. 
Yet, these variations are rather moderate and we can well estimate the overall header size as $0.5$ kB.
Keeping these quantities in mind, we can make some considerations about the required storage capacities.
Clearly, running a full node is something that becomes challenging even for a standard personal computer. Indeed, syncing to the chain means that every block must be downloaded and stored: this would require 626 GB.
We argue that even running a light node may require a significant memory cost. 
Indeed, to store all the block headers, we need approximately
$14,497,082\cdot 0.5\text{ kB} \approx 7.2 \text{ GB}$, which is clearly way more than the storage capacity of a standard IoT device.
Furthermore, given that Ethereum produces a block every 13.2 seconds, this means that every day approximately 6,545 blocks are produced, yielding to more than 3 MB to store the corresponding new headers.

This simple analysis shows that the functionalities of both full and light nodes are out of range for typical IoT devices.
Accounts, on the contrary, do not need to store anything about the blockchain, but simply need to somehow have access to their personal data (key-pair and address) and keep track of the sent transactions (that is, the field \texttt{nonce} in the transaction body).
The nonce has size 8 B, while the key-pair requires no more than 96 B (32 B for the secret key, and 64 B for the public key) and the address size is 20 B. This gives a total of 124 B. Note that this dimension can be reduced if we consider that the public key can be computed from the secret key, and the same can be done for the address.
Hence, an account may need to store only 40 B of data.
For these reasons, we argue that the simplest (and, perhaps, the only feasible) way to make an IoT device interact with a blockchain network is that of having it behaving as an account.
This way, it only needs to generate, sign and format transactions, and does not actively participate in the consensus mechanism.
The constructed transactions will then be sent to more powerful devices running a node (light or full), which will forward them to the network by including them into blocks and taking charge of the block construction and proposal.

Note that, to instantiate an Ethereum-compliant account, the following steps need to be performed:
\begin{enumerate}
\item [a) ]generate a secret ECDSA key;
\item[b) ] compute the corresponding ECDSA public key;
\item[c) ] hash the public key with Keccak256 and select the last 20 bytes to obtain the corresponding address.
\end{enumerate}
Once one possesses an account, to prepare and send a transaction, the following algorithmic steps should be performed:
\begin{enumerate}
\item fill up all the transaction fields with consistent data; to do this, set $(\texttt{r},\texttt{s},\texttt{v})$ as $(0,0,\texttt{ChainID})$;
\item encode all the data of the previous step through the RLP encoding; 
\item hash the transaction given as output of the previous step using Keccak256;
\item sign the hashed transaction through ECDSA, obtaining some values for the $(\texttt{r},\texttt{s},\texttt{v})$ parameters;
\item encode all the data of the transaction including the new values for $(\texttt{r},\texttt{s},\texttt{v})$ obtained in the previous step through RLP encoding, in order to obtain the so-called \textit{raw transaction}. 
\end{enumerate}
The result of these steps is an Ethereum-compliant transaction that is ready to be sent and accepted by the network.

\subsection{Security guarantees}

Consider a setting in which a device, holding a secret key \texttt{sk} associated to a public \texttt{address}, performs all the steps we have described in the previous section and prepares signed transactions. 
These transactions are submitted to the network, get validated and consequently included in blocks.
This simple framework can provide data integrity and authenticity, since:
\begin{itemize}
\item[-] data included in transactions cannot be manipulated. Indeed, since \texttt{sk} is kept secret, a modification in the field \texttt{data} would also imply a change in the signature (that is, the values $\texttt{r, s, v}$). However, this is not possible without knowing the secret key;
\item[-] the address can be derived from the signing parameters ($\texttt{r,s,v}$). Hence, signature verification additionally provides guarantees about the source of the data.
\end{itemize}

\section{Validation and Performance Assessment}
\label{performances}
To test the feasibility of running accounts and generating transactions on an embedded system, we have used as hardware platform a STM32 Nucleo-144 developing board from STMicroelectronics. It is based on an STM32F439ZI microcontroller with a 32-bit ARM Cortex-M4 core that operates at a frequency of 180MHz. The board includes 144 multi-function pins and an ST-LINK debugger that allows programming it via USB. The device is equipped with a 256 kB static random access memory and a 2 MB Flash memory, inside which the executable code is stored. The board gives an optimized implementation of cryptographic algorithms, such as some symmetric-key encryption algorithms and different hashing functions. The device is equipped with various peripherals, and a random number generator (RNG). 

Regarding the software setting, we have used STM32CubeIDE \cite{stm32ide} for writing and executing code on the embedded system. Indeed, this tool is a C/C++ development and debugging IDE (Integrated Development Environment) for STM32 microcontrollers and microprocessors. It is used to easily configure peripherals and to generate and compile code. 
For our aims, we have used USART3 and RNG peripherals. The USART3 module allows synchronous and asynchronous serial communication between different devices. This peripheral was used to exchange information between the microcontroller and the computer for data extraction and analysis. 
Finally, we have used the on-board RNG for generating a 32 B random number to be used as a secret key.

Besides the default libraries (such as those that are characteristic of ARM microprocessors), we have used the following external libraries:
\begin{itemize}
\item Micro-ECC \cite{micro-ecc} - provides a small and fast ECDSA implementation for microprocessors. This library supports different curves, including the \textit{secp256k1} curve. 
\item Ethereum-RLP \cite{rlp} - provides a C implementation of the Ethereum RLP encoding for the serialization of transactions. 
\item RHash \cite{rhash}, which we have used to implement the hash function Keccak256. It is needed both for generating the address from the public key and for computing the digest of the transaction after serialization. 
\end{itemize}

\subsection{Benchmarking ECDSA}

\begin{figure} [b]
\centering
\includegraphics[width=\columnwidth]{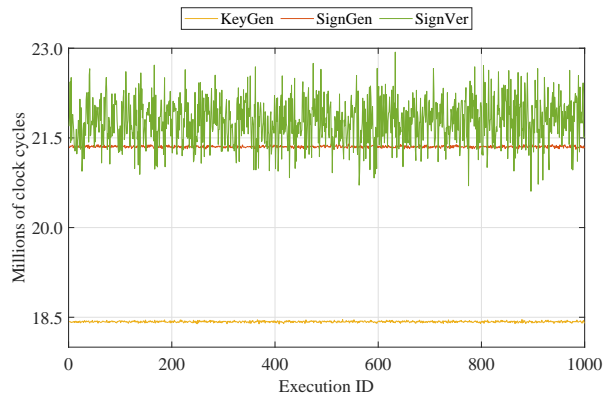}
    \caption{Performance, in terms of clock cycles, of the ECDSA functions (key generation, signature generation, and signature verification) for the $1000$ random test executions.}
    \label{fig: all_rr_xcube}
\end{figure}

\begin{figure} [ht]
\centering
\includegraphics[width=\columnwidth]{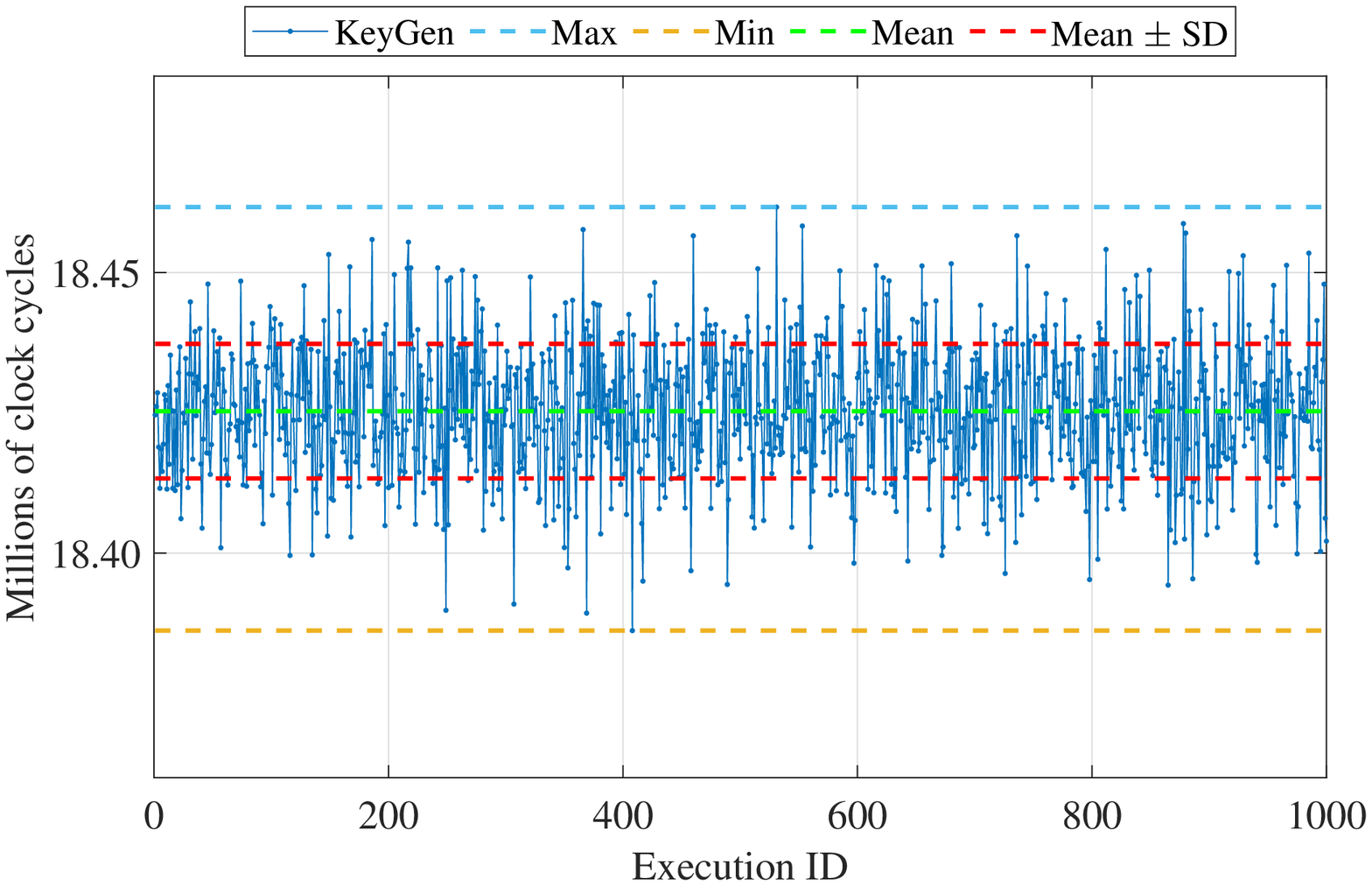}
    \caption{Performance of the key generation function in terms of clock cycles for $1000$ random tests. The minimum, the maximum, and the mean values are highlighted, along with the standard deviation (SD) values.}
    \label{fig: keygen_rr_xcube}
\end{figure}

\begin{figure} [ht]
\centering
\includegraphics[width=\columnwidth]{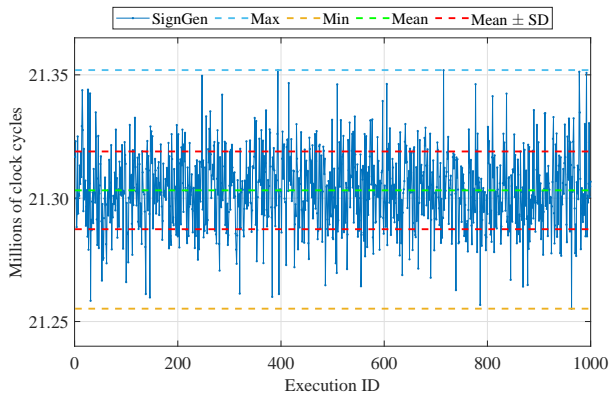}
    \caption{Performance of the signature generation function in terms of clock cycles for $1000$ random tests. The minimum, the maximum, and the mean values are highlighted, along with the standard deviation (SD) values.}
    \label{fig: signgen_rr_xcube}
\end{figure}

We first report some performance measures concerning ECDSA, which we expect to represent the most burdensome function from a computational point of view.
In order to provide a complete picture of the considered ECDSA implementation, we have chosen to also show the performance of the signature verification, even though this algorithm is normally not used by an account.
We have executed the code to create $1,000$ randomly generated raw transactions by drawing, for each execution, a random private key and random $\texttt{data}$.
Figure \ref{fig: all_rr_xcube} shows a performance comparison of the three ECDSA algorithms, measured in terms of millions of clock cycles. 
The performance of key generation alone is reported in Figure \ref{fig: keygen_rr_xcube}, while Figure \ref{fig: signgen_rr_xcube} shows the benchmarks for signature generation. 

We see that the key and signature generation functions have always taken no more than $21.5$ millions of clock cycles, resulting in an expected time lower than $\frac{21.5\cdot 10^6 }{180\text{ MHz}}\approx 0.12$ s (assuming that one clock cycle takes $\frac{1}{180\text{ MHz}}$ seconds).
These results confirm the feasibility of running the ECDSA algorithm on the considered platform.
Table \ref{table:confronto} reports, for each function, the average running time, the observed maximum and minimum values and the standard deviation (referred to as SD).
The results are expressed, again, in terms of millions of clock cycles. 
As another important observation, we note that the measured standard deviation values for the KeyGen and SignGen algorithms are much smaller than the average value (by three orders of magnitude). 
This behaviour is also confirmed by Figure \ref{fig: all_rr_xcube}, since the fluctuations of KeyGen and SignGen are rather limited.
This shows that the considered implementation offers some level of protection against side-channel attacks. 
In fact, the standard deviation of an algorithm running time is an aggregate measure of the fluctuations as a function of the input features: the lower the standard deviation, the smaller the amount of information which can be leaked by the execution of the algorithm.

\begin{table}[ht]
\begin{center}
\caption{Performance of the Micro-ECC library implementing ECDSA functions, in terms of millions of clock cycles.}
\begin{tabular}{|c|c|c|c|c|}
\hline
Function                           & Min  & Max  & Mean  & SD \\ \hline\hline
KeyGen & $18.38$ & $18.46$ & $18.42$ & $0.01$ \\ \hline
SignGen & $21.29$ & $21.40$ & $21.34$ & $0.02$ \\ \hline
SignVer & $20.33$ & $22.76$ & $21.66$ & $0.42$ \\\hline
\end{tabular}
\label{table:confronto}
\end{center}
\end{table}

\subsection{Benchmarking the whole transaction generation}

In this section, we report the timing results for the full generation of an Ethereum-compliant transaction.
For each generated transaction, we have measured the number of cycles required to perform each of the steps 2-5 reported in Section \ref{thecost}.
The results are reported in Table \ref{tab:fulltx}, where we have also included the estimated amount of time required for each operation, as well as the time percentage.
For all the considered operations, we have observed a rather small difference between the minimum and maximum values, so that here we only report the mean values.
\begin{table} [ht!]
\centering
 \caption{Number of required clock cycles and amount of time for each phase of the transaction creation algorithm.}
\begin{tabular}{ |c|cc|c| }
 \hline
Phase & Average number of cycles & Average time (ms) & Perc. \\ 
\hline\hline
$2$ & $2.7 \cdot 10^4$ & $0.15$ & $0.12\%$ \\
\hline
$3$ & $6.0 \cdot 10^5$ & $3.33$ & $2.77\%$\\
\hline
$4$ & $2.1 \cdot 10^7$ & $116.67$ & $96.88\%$ \\
\hline
$5$ & $5.0 \cdot 10^4$ & $0.28$ & $0.23\%$ \\
\hline
 \end{tabular}
\label{tab:fulltx}
\end{table}
By summing all the values in the last-but-one column of the table, we see that the generation of a transaction requires roughly $120$ ms, on average.
As we expected, the majority of this time is spent for the signature generation, which takes approximately $97\%$ of the total time.
All of the other functions require a negligible amount of time, apart from the hash computation (Phase 3) that, however, takes only $2.77\%$ of the total execution time.

Notice that, to obtain these measurements, we have excluded the timings of the ECDSA key generation algorithm, since it can be run just once if there is enough memory to store the key-pair and the corresponding address.
If one desires to save some memory, a possibility is that of keeping only the seed associated to the secret key so that, when producing a new transaction, it is enough to use it to re-generate the key pair and the account address.
With this approach, generating and sending a transaction would require approximately $222$ ms (since the KeyGen algorithm takes on average $102$ ms).

\subsection{Benchmarking the complete framework}

The results in the previous section show that the considered device can efficiently handle all the operations which are required to construct valid transactions that are fully compliant with the Ethereum blockchain.
To validate the considered procedure, we have run tests using the Ropsten testnet. Namely, we have run an experiment in which the STM board prepares transactions and sends them to Ropsten. 
We have considered the following procedure:
\begin{enumerate}
\item the STM board asks the blockchain for the nonce of the last transaction from its address. The received value is used as the initial nonce;
\item the board starts producing transactions, which are then submitted to the blockchain;
\item after sending a new transaction, the board waits for an acknowledge;
\item if the acknowledge is properly received, the board starts producing new transactions. 
\end{enumerate}
Communication with the blockchain is necessary to guarantee that the board is correctly synchronized with the network. Indeed, a transaction will be discarded if its nonce is too low, or will be pending in case if too high.
For instance, if the board sends a transaction having an insufficient balance, this will be deemed as invalid.
In this case, the board should i) wait for its balance to be refilled, and ii) reuse the nonce associated to the invalid transaction.
The acknowledge received by the network guarantees that the board stops sending new transactions, until potential problems have been solved. This step has been introduced to avoid the situation in which the board keeps producing transactions (and, so, locally increases its nonce) but works in a non-synchronous way with respect to the blockchain evolution.
These issues arise since the considered device operates as an account, without downloading and storing the whole blockchain. Indeed, running a full node, the correct nonce could be easily recovered by scanning blocks.

To run a full test for the proposed infrastructure, we have simulated the interaction between the board and the blockchain using a serial communication with a laptop.
In our experiments, we have used a laptop with an Intel(R) Core(TM) i7-8565U CPU processor, with a 1.80GHz frequency.  
The computer has a 16 GB RAM and uses Windows 11.
The communication between the laptop and the board has been handled by a Python script, using the standard serial communication with baud rate 115,200.
Furthermore, the Python script has been used to interact with the Ropsten testnet, exploiting the dedicated Web3 API.
The script we have used for the experiments, together with the C code for the STM microcontroller, are publicly available\footnote{\url{https://github.com/secomms/CortexM4Ethereum}}.
We notice that the laptop is only responsible for the communication between the board and the blockchain, and cannot manipulate transactions (since the secret key is held by the STM board). Therefore, data integrity and authenticity is guaranteed.

We have repeated the tests for several accounts sending at least 100 transactions for each one.
Averaging over all trials and measuring the required times, we found that:
\begin{itemize}
\item the laptop receives a new transaction, on average, every  $0.259$ ms;
\item the time to send the transaction to the Ropsten testnet and receive the receipt is, on average, $0.215$ ms.
\end{itemize}
Hence, on average, the considered setup can send a new transaction to the Ropsten network in  $\approx 0.474$ ms.

\section{Conclusion}
\label{conclusion}
We have proposed and studied an implementation of the functions needed for realizing an Ethereum account and Ethereum-compliant transactions on an STM32 microcontroller with an ARM Cortex-M4 microprocessor, which is a typical hardware architecture for IoT devices. 
We have shown that the best and most feasible way for integrating IoT and blockchain is that an IoT device acts as an account, sending prepackaged and immutable transactions containing relevant information, such as data collected from sensors, to an overlying blockchain-connected node.
We have implemented all the functions required for creating Ethereum-compliant transactions on an STM32 microcontroller and we have assessed the performance achievable by our implementation, validating it on the Ropsten testnet. The benchmark results show that the approach we propose is actually feasible and compatible with the needs of IoT devices and infrastructures in terms of both efficiency and performance.

\section*{Acknowledgments}
The authors wish to thank Domenico Andrea Giliberti for his precious help in the implementation and benchmarking of the algorithms which have been considered in this paper.

\bibliographystyle{IEEEtran}
\bibliography{ref}

\begin{thebibliography}{10}
\providecommand{\url}[1]{#1}
\csname url@samestyle\endcsname
\providecommand{\newblock}{\relax}
\providecommand{\bibinfo}[2]{#2}
\providecommand{\BIBentrySTDinterwordspacing}{\spaceskip=0pt\relax}
\providecommand{\BIBentryALTinterwordstretchfactor}{4}
\providecommand{\BIBentryALTinterwordspacing}{\spaceskip=\fontdimen2\font plus
\BIBentryALTinterwordstretchfactor\fontdimen3\font minus
  \fontdimen4\font\relax}
\providecommand{\BIBforeignlanguage}[2]{{%
\expandafter\ifx\csname l@#1\endcsname\relax
\typeout{** WARNING: IEEEtran.bst: No hyphenation pattern has been}%
\typeout{** loaded for the language `#1'. Using the pattern for}%
\typeout{** the default language instead.}%
\else
\language=\csname l@#1\endcsname
\fi
#2}}
\providecommand{\BIBdecl}{\relax}
\BIBdecl

\bibitem{madakam}
S.~Madakam, R.~Ramaswamy, and S.~Tripathi, ``{I}nternet of things ({I}o{T}): a
  literature review,'' \emph{Journal of Computer and Communications}, vol.~3,
  no.~05, pp. 164--173, 2015.

\bibitem{xiao}
P.~Xiao, \emph{Designing embedded systems and the {I}nternet of {T}hings (IoT)
  with the {ARM} mbed}.\hskip 1em plus 0.5em minus 0.4em\relax John Wiley \&
  Sons, 2018.

\bibitem{diaz}
M.~D{\'\i}az, C.~Mart{\'\i}n, and B.~Rubio, ``State-of-the-art, challenges, and
  open issues in the integration of {I}nternet of things and cloud computing,''
  \emph{J. Netw. Comput. Appl.}, vol.~67, pp. 99--117, 2016.

\bibitem{malviya}
H.~Malviya, ``How blockchain will defend {I}o{T},'' \emph{Available at SSRN
  2883711}, 2016.

\bibitem{yiu}
J.~Yiu and A.~Frame, ``Cortex-{M} processors and the {I}nternet of {T}hings
  ({IoT}),'' \emph{White paper}, January 2013.

\bibitem{reyna}
A.~Reyna, C.~Mart{\'\i}n, J.~Chen, E.~Soler, and M.~D{\'\i}az, ``On blockchain
  and its integration with {I}o{T}. {C}hallenges and opportunities,''
  \emph{Future generation computer systems}, vol.~88, pp. 173--190, 2018.

\bibitem{conoscenti}
M.~Conoscenti, A.~Vetrò, and J.~C. De~Martin, ``Blockchain for the internet of
  things: A systematic literature review,'' in \emph{2016 IEEE/ACS 13th
  International Conference of Computer Systems and Applications (AICCSA)},
  2016, pp. 1--6.

\bibitem{ledwaba}
L.~P. Ledwaba, G.~P. Hancke, H.~S. Venter, and S.~J. Isaac, ``Performance costs
  of software cryptography in securing new-generation internet of energy
  endpoint devices,'' \emph{IEEE Access}, vol.~6, pp. 9303--9323, 2018.

\bibitem{umran}
S.~M. Umran, S.~Lu, Z.~A. Abduljabbar, J.~Zhu, and J.~Wu, ``Secure data of
  industrial internet of things in a cement factory based on a {B}lockchain
  technology,'' \emph{Applied Sciences}, vol.~11, no.~14, p. 6376, 2021.

\bibitem{latif}
S.~Latif, Z.~Idrees, J.~Ahmad, L.~Zheng, and Z.~Zou, ``A blockchain-based
  architecture for secure and trustworthy operations in the industrial
  {I}nternet of {T}hings,'' \emph{Journal of Industrial Information
  Integration}, vol.~21, p. 100190, 2021.

\bibitem{firefly}
\BIBentryALTinterwordspacing
``A simple, low-cost {E}thereum wallet,'' accessed: 04-2022. [Online].
  Available: \url{https://github.com/firefly/wallet}
\BIBentrySTDinterwordspacing

\bibitem{ethereum}
G.~Wood, ``Ethereum: A secure decentralised generalised transaction ledger,''
  \emph{Ethereum project yellow paper}, 2021.

\bibitem{ecdsa}
D.~Johnson, A.~Menezes, and S.~Vanstone, ``The elliptic curve digital signature
  algorithm (ecdsa),'' \emph{International journal of information security},
  vol.~1, no.~1, pp. 36--63, 2001.

\bibitem{eip155}
V.~Buterin, ``{EIP}-155: {S}imple replay attack protection,'' \emph{Ethereum
  Improvement Proposals}, no. 155, October 2016.

\bibitem{stm32ide}
\BIBentryALTinterwordspacing
ST, ``Integrated development environment for {STM32}.'' [Online]. Available:
  \url{https://www.st.com/stm32cubeide}
\BIBentrySTDinterwordspacing

\bibitem{micro-ecc}
\BIBentryALTinterwordspacing
``A small and fast {ECDH} and {ECDSA} implementation for 8-bit, 32-bit, and
  64-bit processors,'' accessed: 04-2022. [Online]. Available:
  \url{https://github.com/kmackay/micro-ecc}
\BIBentrySTDinterwordspacing

\bibitem{rlp}
\BIBentryALTinterwordspacing
``C {E}thereum {RLP} library,'' accessed: 04-2022. [Online]. Available:
  \url{https://github.com/KingHodor/Ethereum-RLP}
\BIBentrySTDinterwordspacing

\bibitem{rhash}
\BIBentryALTinterwordspacing
``Rhash ({R}ecursive {H}asher),'' accessed: 04-2022. [Online]. Available:
  \url{https://github.com/rhash/RHash/blob/master/librhash/sha3.c}
\BIBentrySTDinterwordspacing

\end{thebibliography}


\begin{thebibliography}{10}
\providecommand{\url}[1]{#1}
\csname url@samestyle\endcsname
\providecommand{\newblock}{\relax}
\providecommand{\bibinfo}[2]{#2}
\providecommand{\BIBentrySTDinterwordspacing}{\spaceskip=0pt\relax}
\providecommand{\BIBentryALTinterwordstretchfactor}{4}
\providecommand{\BIBentryALTinterwordspacing}{\spaceskip=\fontdimen2\font plus
\BIBentryALTinterwordstretchfactor\fontdimen3\font minus
  \fontdimen4\font\relax}
\providecommand{\BIBforeignlanguage}[2]{{%
\expandafter\ifx\csname l@#1\endcsname\relax
\typeout{** WARNING: IEEEtran.bst: No hyphenation pattern has been}%
\typeout{** loaded for the language `#1'. Using the pattern for}%
\typeout{** the default language instead.}%
\else
\language=\csname l@#1\endcsname
\fi
#2}}
\providecommand{\BIBdecl}{\relax}
\BIBdecl

\bibitem{iot_review}
S.~Madakam, R.~Ramaswamy, and S.~Tripathi, ``{I}nternet of things ({I}o{T}): a
  literature review,'' \emph{Journal of Computer and Communications}, vol.~3,
  no.~05, pp. 164--173, 2015.

\bibitem{perry}
P.~Xiao, \emph{Designing embedded systems and the {I}nternet of {T}hings (IoT)
  with the {ARM} mbed}.\hskip 1em plus 0.5em minus 0.4em\relax John Wiley \&
  Sons, 2018.

\bibitem{iot_cloud}
M.~D{\'\i}az, C.~Mart{\'\i}n, and B.~Rubio, ``State-of-the-art, challenges, and
  open issues in the integration of {I}nternet of things and cloud computing,''
  \emph{J. Netw. Comput. Appl.}, vol.~67, pp. 99--117, 2016.

\bibitem{iot_blockchain}
H.~Malviya, ``How blockchain will defend {I}o{T},'' \emph{Available at SSRN
  2883711}, 2016.

\bibitem{arm_iot}
J.~Yiu and A.~Frame, ``Cortex-{M} processors and the {I}nternet of {T}hings
  ({IoT}),'' \emph{White paper}, January 2013.

\bibitem{reyna}
A.~Reyna, C.~Mart{\'\i}n, J.~Chen, E.~Soler, and M.~D{\'\i}az, ``On blockchain
  and its integration with {I}o{T}. {C}hallenges and opportunities,''
  \emph{Future generation computer systems}, vol.~88, pp. 173--190, 2018.

\bibitem{conoscenti}
M.~Conoscenti, A.~Vetrò, and J.~C. De~Martin, ``Blockchain for the internet of
  things: A systematic literature review,'' in \emph{2016 IEEE/ACS 13th
  International Conference of Computer Systems and Applications (AICCSA)},
  2016, pp. 1--6.

\bibitem{performance}
L.~P. Ledwaba, G.~P. Hancke, H.~S. Venter, and S.~J. Isaac, ``Performance costs
  of software cryptography in securing new-generation internet of energy
  endpoint devices,'' \emph{IEEE Access}, vol.~6, pp. 9303--9323, 2018.

\bibitem{cement}
S.~M. Umran, S.~Lu, Z.~A. Abduljabbar, J.~Zhu, and J.~Wu, ``Secure data of
  industrial internet of things in a cement factory based on a {B}lockchain
  technology,'' \emph{Applied Sciences}, vol.~11, no.~14, p. 6376, 2021.

\bibitem{latif}
S.~Latif, Z.~Idrees, J.~Ahmad, L.~Zheng, and Z.~Zou, ``A blockchain-based
  architecture for secure and trustworthy operations in the industrial
  {I}nternet of {T}hings,'' \emph{Journal of Industrial Information
  Integration}, vol.~21, p. 100190, 2021.

\bibitem{firefly}
\BIBentryALTinterwordspacing
``A simple, low-cost {E}thereum wallet,'' accessed: 04-2022. [Online].
  Available: \url{https://github.com/firefly/wallet}
\BIBentrySTDinterwordspacing

\bibitem{ethereum}
G.~Wood, ``Ethereum: A secure decentralised generalised transaction ledger,''
  \emph{Ethereum project yellow paper}, 2021.

\bibitem{ecdsa}
D.~Johnson, A.~Menezes, and S.~Vanstone, ``The elliptic curve digital signature
  algorithm (ecdsa),'' \emph{International journal of information security},
  vol.~1, no.~1, pp. 36--63, 2001.

\bibitem{eip155}
V.~Buterin, ``{EIP}-155: {S}imple replay attack protection,'' \emph{Ethereum
  Improvement Proposals}, no. 155, October 2016.

\bibitem{stm32ide}
\BIBentryALTinterwordspacing
ST, ``Integrated development environment for {STM32}.'' [Online]. Available:
  \url{https://www.st.com/stm32cubeide}
\BIBentrySTDinterwordspacing

\bibitem{micro-ecc}
\BIBentryALTinterwordspacing
``A small and fast {ECDH} and {ECDSA} implementation for 8-bit, 32-bit, and
  64-bit processors,'' accessed: 04-2022. [Online]. Available:
  \url{https://github.com/kmackay/micro-ecc}
\BIBentrySTDinterwordspacing

\bibitem{rlp}
\BIBentryALTinterwordspacing
``C {E}thereum {RLP} library,'' accessed: 04-2022. [Online]. Available:
  \url{https://github.com/KingHodor/Ethereum-RLP}
\BIBentrySTDinterwordspacing

\bibitem{rhash}
\BIBentryALTinterwordspacing
``Rhash ({R}ecursive {H}asher),'' accessed: 04-2022. [Online]. Available:
  \url{https://github.com/rhash/RHash/blob/master/librhash/sha3.c}
\BIBentrySTDinterwordspacing

\end{thebibliography}

\end{document}